\documentclass{elsart}

\usepackage{natbib}

\usepackage{amssymb}

\begin{document}

\begin{frontmatter}



\title{Generalized Fokker-Planck equations \\
and effective thermodynamics}


\author{Pierre-Henri Chavanis}

\address{Laboratoire de Physique Th\'eorique, Universit\'e
Paul Sabatier\\ 118 route de Narbonne, 31062 Toulouse Cedex 4,
France\\
E-mail: chavanis@irsamc.ups-tlse.fr}

\begin{abstract}
We introduce a new class of Fokker-Planck equations associated with an
effective generalized thermodynamical framework. These equations
describe a gas of Langevin particles in interaction. The free energy
can take various forms which can account for anomalous diffusion,
quantum statistics, lattice models... When the potential of
interaction is long-ranged, these equations display a rich structure
associated with canonical phase transitions and blow-up phenomena. In
the limit of short-range interactions, they reduce to Cahn-Hilliard
equations.
\end{abstract}


\end{frontmatter}

% main text

\section{Introduction}
\label{intro}

The statistical mechanics of systems with long-range interactions is
currently a topic of active research. So far, most works have focused
on the case of Hamiltonian systems of particles such as the $N$-stars
problem, the $N$-vortex problem or the HMF model [1]. For
these systems the energy is conserved and the proper statistical
description is the microcanonical ensemble. The statistical evolution
of the particles is described by kinetic equations such as the Landau
equation in astrophysics [2] or the kinetic equation
derived in [3] for point vortices. Since statistical ensembles
are not equivalent for systems with long-range interactions, it may be
of interest, at a conceptual level, to compare this microcanonical
evolution with a canonical one. In that respect, we can consider a
system of Brownian particles in interaction. These particles are in
contact with a thermal bath that imposes its temperature. Therefore,
they have a rigorous canonical structure and their statistical
evolution is described by non-local Fokker-Planck equations [4,5].

In a different context, it has been shown that Tsallis generalized
thermodynamics could be useful to interpret anomalous diffusion in
complex systems and that the $q$-entropies are connected to nonlinear
Fokker-Planck equations [6,7]. These equations can be derived from
stochastic processes involving a multiplicative noise depending on the
distribution function [8]. In fact, it is possible to construct a
larger class of Fokker-Planck equations associated with a generalized
thermodynamical framework encompassing Tsallis and Boltzmann entropies
[9-12,5]. These generalized Fokker-Planck equations are obtained from
standard ones by assuming that the diffusion coefficient is an
arbitrary function $D(f)$ of the distribution function (the friction
term can also depend on $f$).  For these equations, there exists a
Lyapunov functional, satisfying $\dot F\le 0$, that can be interpreted
as a generalized free energy. In this context, Tsallis free energy
corresponds to a power-law dependence of the diffusion coefficient
$D(f)\sim f^{q-1}$ and the $q$ parameter in Tsallis formalism is
related to the exponent of anomalous diffusion.

In this paper, following [11], we introduce a new class of
Fokker-Planck equations associated with a generalized free energy
functional {\it and} an arbitrary potential of interaction. These
equations connect the two topics discussed above: long-range
interactions and generalized stochastic processes. Their mathematical
and physical richness is considerable, as they exhibit a wide variety
of canonical phase transitions and blow-up phenomena [13-17]. In the
limit of short-range interactions, they reduce to Cahn-Hilliard
equations.

\section{Generalized Fokker-Planck equations in phase space}
\label{ps}

\subsection{Langevin particles in an external potential}
\label{pse}

We consider a system of non-interacting Langevin particles evolving in
a fixed external potential $\Phi_{ext}({\bf r})$. We assume that the
motion of each particle is described by the generalized stochastic
process
\begin{eqnarray}
\label{pse1}
{d{\bf r}\over dt}={\bf v},\quad
{d{\bf v}\over dt}=-\xi{\bf v}-\nabla\Phi_{ext}({\bf r})+\sqrt{2Df\biggl\lbrack {C(f)\over f}\biggr\rbrack'}{\bf R}(t),
\end{eqnarray}
where ${\bf R}(t)$ is a white noise
satisfying $\langle {\bf R}(t)\rangle ={\bf 0}$ and $\langle
R_a(t)R_b(t')\rangle=\delta_{ab}\delta(t-t')$ and $C$ is an arbitrary
convex function, i.e. $C''>0$. Since the function in front of ${\bf
R}(t)$ depends on $({\bf r},{\bf v})$, the last term in
Eq. (\ref{pse1}) can be interpreted as a multiplicative noise. We
note, however, that it depends on $({\bf r},{\bf v})$ through the
local density of particles $f({\bf r},{\bf v},t)$. This corresponds to
a back-reaction from the macroscopic dynamics. When $C(f)={1\over
q-1}(f^{q}-f)$, Eq. (\ref{pse1}) reduces to the
stochastic process studied by Borland [8] in connexion with
Tsallis thermodynamics.

Using standard methods, we can show that the stochastic process
(\ref{pse1}) leads to the generalized Fokker-Planck equation
\begin{equation}
\label{pse2} {\partial f\over\partial t}+{\bf v}\cdot {\partial
f\over\partial {\bf r}}-\nabla\Phi_{ext}\cdot {\partial f\over\partial
{\bf v}}={\partial \over \partial {\bf v}}\cdot \biggl\lbrace
D\biggl\lbrack f C''(f){\partial f\over\partial {\bf v}}+\beta
f{\bf v}\biggr \rbrack \biggr\rbrace,
\end{equation}
where we have introduced a generalized inverse temperature $\beta=1/T$
through the generalized Einstein relation $\xi=D\beta$.  This
equation can be put in the form
\begin{equation}
\label{pse3} {\partial f\over\partial t}+Lf ={\partial\over\partial
{\bf v}}\cdot \biggl \lbrack \xi f {\partial\over\partial {\bf v}}\biggl
({\delta F\over\delta f}\biggr )\biggr\rbrack,
\end{equation}
where $L$ is the advection operator and $F$ is the generalized free energy
\begin{equation}
\label{pse4} F[f]=E-TS=\int f {v^2\over 2}d^{D}{\bf r}d^{D}{\bf v}+
\int \rho \Phi_{ext}d^{D}{\bf r}+T\int C(f)d^{D}{\bf r}d^{D}{\bf
v}.
\end{equation}
The first term in Eq. (\ref{pse4}) is the kinetic energy, the second
term is the potential energy and the third term is a generalized
entropy $S=-\int C(f)d^{D}{\bf r}d^{D}{\bf v}$. It is easy to show
that the dissipation of free energy can be expressed as
\begin{equation}
\label{pse5} \dot F=-\int {DT\over f}\biggl\lbrack fC''(f){\partial
f\over\partial {\bf v}}+\beta f {\bf v}\biggr \rbrack^{2}
d^{D}{\bf r}d^{D}{\bf v},
\end{equation}
which is negative provided that $D>0$. Therefore, the generalized
Fokker-Planck equation (\ref{pse2}) satisfies a canonical H-theorem $\dot
F\le 0$. Hence, the free energy (\ref{pse4}) is a Lyapunov functional.
Finally, the stationary solutions of Eq. (\ref{pse2}) are given by
\begin{equation}
\label{pse6}
C'(f)=-\beta\biggl ({v^{2}\over 2}+\Phi_{ext}\biggr )-\alpha.
\end{equation}
They extremize the free energy (\ref{pse4}) at fixed mass and
temperature.  In addition, only minima of free energy are linearly
stable via Eq. (\ref{pse2}), see [11].

\subsection{Langevin particles in interaction}
\label{psi}

We now consider a system of Langevin particles in interaction
described by the $N$-body stochastic equations
\begin{eqnarray}
\label{psi1}
{d{\bf r}_{i}\over dt}={\bf v}_{i},\quad
{d{\bf v}_{i}\over dt}=-\xi{\bf v}_{i}-\nabla_{i}U({\bf r}_{1},...,{\bf r}_{N})+\sqrt{2Df_{i}\biggl\lbrack {C(f_{i})\over f_{i}}\biggr\rbrack'}{\bf R}_{i}(t),
\end{eqnarray}
where $f_i=f({\bf r}_i,{\bf v}_i,t)$ and ${\bf R}_i(t)$ is a white
noise acting independently on the particles. The particles
interact via the potential $U=\sum_{i<j}u({\bf r}_i-{\bf r}_j)$
where  $u({\bf r}_i-{\bf r}_j)$ is an arbitrary binary potential
depending only on the absolute distance between particles. When
$C(f)=-f\ln f$, corresponding to the Boltzmann entropy,  Eq. (\ref{psi1})
describes a system of Brownian particles in interaction.

Starting from the $N$-body Fokker-Planck equation, using a
Kramers-Moyal expansion and a mean-field approximation
[4,5], we can derive from (\ref{psi1}) the non-local
generalized Kramers equation
\begin{equation}
\label{psi2} {\partial f\over\partial t}+{\bf v}\cdot {\partial
f\over\partial {\bf r}}-\nabla\Phi\cdot {\partial f\over\partial {\bf
v}}={\partial \over
\partial {\bf v}}\cdot \biggl\lbrace D\biggl\lbrack f C''(f){\partial
f\over\partial {\bf v}}+\beta f{\bf v}\biggr \rbrack
\biggr\rbrace,
\end{equation}
where $\Phi$ is related to the density $\rho=\int f d^{D}{\bf v}$ by a relation of the
form
\begin{equation}
\label{psi3}\Phi({\bf r},t)=\int \rho({\bf r}',t)u({\bf r}-{\bf
r'})d^{D}{\bf r}'.
\end{equation}
Equation (\ref{psi2}) can be written as Eq. (\ref{pse3}) where, now,  the free
energy is given by
\begin{equation}
\label{psi4} F[f]=\int f {v^2\over 2}d^{D}{\bf r}d^{D}{\bf v}+{1\over
2}\int \rho \Phi d^{D}{\bf r}+T\int C(f)d^{D}{\bf r}d^{D}{\bf v}.
\end{equation}
Furthermore, Eq. (\ref{pse5}) remains valid so that $\dot F\le 0$. Finally,
the stationary states of Eq. (\ref{psi2}) are determined by the
integro-differential equation
\begin{equation}
\label{psi5} C'(f)=-\beta\biggl\lbrace {v^{2}\over 2}+\int f({\bf r}',{\bf v}')u({\bf r}-{\bf r}')d^{D}{\bf r}'d^{D}{\bf v}'\biggr\rbrace-\alpha.
\end{equation}
They extremize the free energy (\ref{psi4}) at fixed mass and temperature.
In addition, only minima are linearly stable via Eq. (\ref{psi2}), see [11].

\subsection{The strong friction limit}
\label{sf}

In the strong friction limit  $\xi\rightarrow
+\infty$, or equivalently for large times
$t\gg\xi^{-1}$, the distribution function is given by
\begin{eqnarray}
\label{sf1}
C'(f)=-\beta\biggl \lbrack {v^{2}\over 2}+\lambda({\bf r},t)\biggr \rbrack+O(\xi^{-1}),
\end{eqnarray}
where $\lambda$ is related to the density $\rho$, using $\rho=\int f
d^D{\bf v}=\rho(\lambda)$. Introducing the pressure $p={1\over D}\int
f v^2 d^D{\bf v}=p(\lambda)$ and eliminating $\lambda$, we find that
the fluid is barotropic in the sense that $p=p(\rho)$, where the
equation of state is entirely specified by the function
$C(f)$. Furthermore, writing the hierarchy of moment equations
[11] or using a formal Chapman-Enskog expansion
[18], we can show that the evolution of the density is
determined by the non-local generalized Smoluchowski equation
\begin{equation}
\label{sf2} {\partial \rho\over\partial t}=\nabla\cdot\biggl\lbrack
{1\over\xi}(\nabla p+\rho\nabla\Phi)\biggr\rbrack,
\end{equation}
which can be written explicitly as
\begin{equation}
\label{sf3}
{\partial \rho\over\partial t}=\nabla\cdot\biggl\lbrace {1\over\xi}\biggl \lbrack p'(\rho)\nabla \rho+\rho\nabla\int u({\bf r}-{\bf r}')\rho({\bf r}',t)d^{D}{\bf r}'\biggr \rbrack\biggr\rbrace.
\end{equation}
This equation can be put in the form
\begin{equation}
\label{sf4} {\partial \rho\over\partial t}= \nabla\cdot \biggl\lbrack
{1\over\xi}\rho\nabla\biggl ({\delta F\over\delta \rho}\biggr
)\biggr\rbrack,
\end{equation}
where $F$ is the free energy
\begin{eqnarray}
{F}[\rho]=\int \rho \int_{0}^{\rho}{p(\rho')\over\rho'^{2}}d\rho'
d^D{\bf r}+{1\over
  2}\int\rho({\bf r},t)u({\bf r}-{\bf r}')\rho({\bf r}',t) d^{D}{\bf r}d^{D}{\bf r}'.
\label{sf5}
\end{eqnarray}
It can be obtained from Eq. (\ref{psi4}) by using the relation
(\ref{sf1}) valid in the strong friction limit $\xi\rightarrow
+\infty$ to express $F[f]$ as a functional of $\rho$ [11,18]. It is
straightforward to show that
\begin{equation}
\label{sf6} \dot F=-\int {1\over \xi\rho} (\nabla
p+\rho\nabla\Phi)^2 d^{D}{\bf r},
\end{equation}
which yields $\dot F\le 0$. Therefore, the free energy (\ref{sf5}) is
a Lyapunov functional for the generalized Smoluchowski equation
(\ref{sf3}). Finally, the stationary solutions of Eq. (\ref{sf3}) are
determined by the condition of hydrostatic equilibrium
\begin{equation}
\label{sf7}\nabla p+\rho\nabla\Phi={\bf 0},
\end{equation}
which can be written explicitly as
\begin{equation}
\label{sf8} {p'(\rho)\over \rho}\nabla \rho=-\nabla\int \rho({\bf
r}')u({\bf r}-{\bf r}')d^{D}{\bf r}'.
\end{equation}
They extremize the free energy (\ref{sf5}) at fixed mass.
In addition, only minima are linearly stable via Eq. (\ref{sf2}).

\section{Generalized Fokker-Planck equations in physical space}
\label{hs}

\subsection{Langevin particles in an external potential}
\label{hse}

We consider a system of non-interacting Langevin particles
described by the generalized stochastic process
\begin{equation}
\label{hse1} {d{\bf r}\over dt}=-\xi\nabla
\Phi_{ext}({\bf r})+\sqrt{2D\rho\biggl\lbrack
{C(\rho)\over\rho}\biggr\rbrack'}{\bf R}(t),
\end{equation}
where $\xi$ is the drift coefficient and ${\bf R}(t)$ is a white
noise. The evolution of the density is described by the
generalized Fokker-Planck equation
\begin{equation}
\label{hse2} {\partial \rho\over\partial t}=\nabla\cdot \biggl\lbrace
D\biggl\lbrack \rho C''(\rho)\nabla\rho+\beta
\rho\nabla\Phi_{ext}\biggr \rbrack \biggr\rbrace,
\end{equation}
where we have introduced a generalized inverse temperature $\beta=1/T$
through the generalized Einstein relation $\xi=D\beta$. This
equation can be put in the form 
\begin{equation}
\label{hse3} {\partial \rho\over\partial t}= \nabla\cdot \biggl\lbrack
{\xi}\rho\nabla\biggl ({\delta F\over\delta \rho}\biggr
)\biggr\rbrack,
\end{equation}
where  $F$ is the generalized free energy
\begin{equation}
\label{hse4} F[\rho]=E-TS=\int \rho \Phi_{ext} d^{D}{\bf r}+T\int
C(\rho)d^{D}{\bf r}.
\end{equation}
The rate of dissipation of free energy can be expressed as
\begin{equation}
\label{hse4b} \dot F=-\int {DT\over \rho}\biggl\lbrack \rho C''(\rho)\nabla\rho+\beta\rho\nabla\Phi_{ext}\biggr \rbrack^{2}d^{D}{\bf r},
\end{equation}
which is negative provided that
$D>0$. Therefore, the generalized Fokker-Planck equation (\ref{hse2})
satisfies a sort of canonical H-theorem $\dot F\le 0$. Hence, the
free energy (\ref{hse4}) is a Lyapunov functional. Finally, the stationary
solutions of Eq. (\ref{hse2}) are given by
\begin{equation}
\label{hse5} C'(\rho)=-\beta\Phi_{ext}-\alpha.
\end{equation}
They extremize the free energy (\ref{hse4}) at fixed mass and
temperature.  In addition, only minima of free energy are linearly
stable via Eq. (\ref{hse2}).

\subsection{Langevin particles in interaction}
\label{hsi}

We now consider a system of Langevin particles in interaction
described by the $N$-body stochastic equations
\begin{equation}
\label{hsi1} {d{\bf r}_{i}\over dt}=-\xi\nabla_{i}U({\bf
r}_{1},...,{\bf r}_{N})+\sqrt{2D\rho_{i}\biggl\lbrack
{C(\rho_{i})\over\rho_{i}}\biggr\rbrack'}{\bf R}_{i}(t),
\end{equation}
where $\rho_i=\rho({\bf r}_i,t)$ and ${\bf R}_i(t)$ is a white
noise. The potential of interaction $U$ is defined as above. Starting
from the $N$-body Fokker-Planck equation, using a Kramers-Moyal
expansion and a mean-field approximation [5], we can derive
from Eq. (\ref{hsi1}) the non-local generalized Smoluchowski equation
\begin{equation}
\label{hsi2} {\partial \rho\over\partial t}=\nabla\cdot \biggl\lbrace
D\biggl\lbrack \rho C''(\rho)\nabla\rho+\beta \rho\nabla\Phi\biggr
\rbrack \biggr\rbrace,
\end{equation}
where $\Phi$ is related to the density $\rho$ by a relation of the
form (\ref{psi3}). Equation (\ref{hsi2}) can be written as (\ref{hse3}) where, now,
the free energy is given by
\begin{equation}
\label{hsi3} F={1\over 2}\int \rho \Phi d^{D}{\bf r}+T\int
C(\rho)d^{D}{\bf r}.
\end{equation}
Furthermore, Eq. (\ref{hse4b}) remains valid so that $\dot F\le 0$. Finally,
the stationary states of Eq. (\ref{hsi2}) are determined by the
integro-differential equation
\begin{equation}
\label{hsi4} C'(\rho)=-\beta\int \rho({\bf r}',t)u({\bf r}-{\bf
r'})d^{D}{\bf r}'-\alpha.
\end{equation}
They extremize the free energy (\ref{hsi3}) at fixed mass and temperature.
In addition, only minima are linearly stable via Eq. (\ref{hsi2}).

\subsection{Connexion with Cahn-Hilliard equations}
\label{ch}

If we now consider the case of short-range interactions, it is
possible to expand the potential
\begin{equation}
\label{ch1}
\Phi({\bf r},t) = \int u({\bf r}')\rho({\bf r}+{\bf r'}) d^D{\bf
  r}',
\end{equation}
in Taylor series for ${\bf r}'\to {\bf 0}$. Introducing the
notations
\begin{equation}
\label{ch2} a=\int u(|{\bf x}|) d^D{\bf x} \;\;\mbox{ and
}\;\; b= {1\over D} \int  u(|{\bf x}|)  x^2 d^D{\bf x}\,,
\end{equation}
we obtain to second order
\begin{equation}
\label{ch3}
\Phi({\bf r},t) = a\rho({\bf r},t) + {b\over 2} \Delta\rho({\bf
  r},t)\,.
\end{equation}
In that limit, the free energy (\ref{hsi3}) takes the form
\begin{equation}
\label{ch4}
F[\rho] = -{b\over 2} \int \left\{ {(\nabla\rho)^2\over 2} +
  V(\rho) \right\} d^D{\bf r}\,,
\end{equation}
where we have set $V(\rho) = -(2T/b)C(\rho) - (a/b)\rho^2$. This is
the usual expression of the Landau free energy. In general $b$ is
negative so we have to minimize this functional integral.  For systems
with short-range interactions, the conservative equation (\ref{hsi2})
becomes
\begin{equation}
\label{ch5} {\partial\rho\over\partial t} = \nabla\cdot
\biggl\lbrace {b\xi\over 2}\rho\nabla \left( \Delta\rho - V'(\rho)
\right) \biggr\rbrace\,.
\end{equation}
This is the Cahn-Hilliard equation which has been extensively studied
in the theory of phase ordering kinetics. Its stationary solutions
describe ``domain walls''. We can view therefore Eq.~(\ref{hsi2})
as a generalization of the Cahn-Hilliard equation to the case of systems 
with long-range interactions.

\section{Conclusion}

In this paper, we have introduced a wide class of generalized
Fokker-Planck equations and we have shown that they were associated
with a consistent thermodynamical formalism. We have derived these
equations from a particular stochastic process involving a special
form of multiplicative noise, but they can also arise from other types
of microscopic models. In our point of view, generalized Fokker-Planck
equations are ``effective equations'' trying to take into account
``hidden constraints'' in complex media that are difficult to
formalize [12]. Since these equations decrease a functional similar to
a generalized free energy, this leads naturally to a notion of
``effective thermodynamics''. This effective approach is not in
contradiction with usual thermodynamics. This is just a heuristic
approach trying to deal with complex situations in which standard
thermodynamics is difficult to implement. We can either use ordinary
thermodynamics (Boltzmann) and try to take into account microscopic
constraints or keep only the most accessible constraints (mass,
energy,...) and introduce an ``effective entropy''. This heuristic
approach may be useful in certain occasions when an exact description
of the system's dynamics is difficult to implement.  Generalized
thermodynamics is justified when all the accessible microstates are
{\it not} equiprobable due to additional microscopic constraints.
These constraints can be due to quantum mechanics (Pauli exclusion
principle), to the existence of a lattice, to hard sphere effects, to
a fractal structure of phase space, etc. We emphasize that our
formalism is valid for more general functionals than the Tsallis
entropy.  Tsallis entropy forms, however, an important class of
functionals associated with polytropic distributions and
power-laws. These distributions generate a natural form of
self-confinement that can be of interest in nonextensive
systems. However, it was our interest here to show that a more general
formalism could be developed consistently. A notion of ``generalized
thermodynamics'', with a different presentation and a different
motivation, has been developed independently by Kaniadakis [9], Frank
[10] and Naudts [19].

The non-local Fokker-Planck equations introduced in this paper can be
of interest in different fields of physics, with various
interpretations. In the usual case where they are associated with the
Boltzmann entropy, they describe a gas of Brownian particles in
interaction [4,5]. Therefore, they constitute the canonical
counterpart of the microcanonical kinetic equations (Boltzmann,
Landau, Lenard-Balescu,...) describing Hamiltonian systems of
particles in interaction [2,3]. They are thus interesting at a
conceptual level to test {\it dynamically} situations of ensemble
inequivalence for systems with long-range interactions. For example,
the ``isothermal collapse'' of self-gravitating Brownian particles
described by the Smoluchowski-Poisson system [13-16] can be compared
with the ``gravothermal catastrophe'' of the Hamiltonian $N$-stars
system described by the Landau-Poisson system (or by the orbit
averaged Fokker-Planck equation). Similarly, the Smoluchowski-Poisson
system for Brownian point vortices can be contrasted with the kinetic
equation obtained in [3] from the Hamiltonian dynamics. They exhibit a
very different behaviour illustrating the inequivalence of statistical
ensembles.

At a practical level, the generalized Fokker-Planck equations
presented in this paper and in [11] can be used as numerical
algorithms to compute explicitly nonlinearly dynamically stable
stationary solutions of the Vlasov and Euler equations. In that
context, the generalized entropy $S=-\int C(f)d^{D}{\bf r}d^{D}{\bf
v}$ is interpreted as a Casimir functional or a $H$-function [20].
Its maximization at fixed energy and mass (or circulation) determines
a condition of dynamical stability, not a condition of thermodynamical
stability. The fact that this variational principle looks similar to
the variational principle that occurs in thermodynamics is essentially
a {\it thermodynamical analogy} [11]. Since the generalized
Fokker-Planck equations introduced in [11] increase the functional $S$
while conserving $E$ and $M$ (or $\Gamma$), they can be used as
physical algorithms to construct arbitrary nonlinearly dynamically
stable stationary solutions of the Vlasov (or Euler) equation. Such
stationary solutions can arise in practice as a result of a (possibly
incomplete) violent relaxation, leading to a metaequilibrium state on
the coarse-grained scale [21,11,12].

Finally, non-local drift-diffusion equations isomorphic to the
Smoluchowski-Poisson system have a real interest in biology in
relation with the concept of chemotaxis (see [17] for a
description of the analogy between self-gravitating Brownian
particles and bacterial populations). Now, it has been noted that the
diffusion of bacteries is anomalous in general so that the diffusion
coefficient which appears in the drift-diffusion equation depends on
the density. This leads to a notion of ``effective'' generalized
thermodynamics. This is simply
because the true dynamics is so complex that it is replaced by a
simple equation with an ``ad hoc'' diffusion coefficient whose
specific form has to be deduced from experiments (this illustrates
what we mean by ``hidden constraints''). Accordingly, different forms of
``generalized free energy'' functionals can arise in biology depending
on the context. 

In conclusion, the generalized Fokker-Planck equations presented in
this paper and in [11] can be of considerable interest, in particular
in the case of long-range interactions where they exhibit a rich
variety of canonical phase transitions. A systematic study of this new
class of equations has been initiated in [13-17].

% \appendix

% \section{}
% \label{}

% Bibliographic references with the natbib package:
% Parenthetical: \citep{Bai92} produces (Bailyn 1992).
% Textual: \citet{Bai95} produces Bailyn et al. (1995).
% An affix and part of a reference:
%   \citep[e.g.][Ch. 2]{Bar76}
%   produces (e.g. Barnes et al. 1976, Ch. 2).


\begin{thebibliography}{}

% \bibitem[Names(Year)]{label} or \bibitem[Names(Year)Long names]{label}.
% (\harvarditem{Name}{Year}{label} is also supported.)


\bibitem{dauxois}  {\small {\it Dynamics and thermodynamics of systems with long range interactions}, edited by Dauxois, T., Ruffo, S., Arimondo, E. and  Wilkens, M. Lecture Notes in Physics, Springer (2002).}

\bibitem{kandrup}  {\small H.E. Kandrup, ApJ {\bf 244}, 316 (1981).}

\bibitem{kin}  {\small P.H. Chavanis, Phys. Rev. E {\bf 64}, 026309 (2001).}

\bibitem{martzel}  {\small N. Martzel and C. Aslangul, J.  Phys. A {\bf 34}, 11225 (2001).}

\bibitem{prep}  {\small  P.H. Chavanis, [cond-mat/0409641].}

\bibitem{plastino}  {\small A.R. Plastino and A. Plastino, Physica A {\bf 222}, 347 (1995) }

\bibitem{bukman}  {\small C. Tsallis and D.J. Bukman, Phys. Rev. E   {\bf 54}, R2197 (1996).}

\bibitem{borland}  {\small L. Borland, Phys. Rev. E {\bf 57}, 6634 (1998)}

\bibitem{kaniadakis}  {\small G. Kaniadakis, Physica A {\bf 296}, 405 (2001).}

\bibitem{frank}  {\small T.D. Frank, Physics Lett. A {\bf 290}, 93 (2001).}

\bibitem{gfp}  {\small P.H. Chavanis, Phys. Rev. E {\bf 68}, 036108 (2003).}

\bibitem{gkt}  {\small P.H. Chavanis, Physica A {\bf 332}, 89 (2004).}

\bibitem{crs}  {\small P.H. Chavanis, C. Rosier and C. Sire, Phys. Rev. E {\bf  66}, 036105 (2002).}

\bibitem{sc1}  {\small  C. Sire and P.H. Chavanis, Phys. Rev. E {\bf  66}, 046133 (2002).}

\bibitem{sc2}  {\small C. Sire and P.H. Chavanis, Phys. Rev. E {\bf 69}, 066109} 

\bibitem{cs}  {\small  P.H. Chavanis and C. Sire, Phys. Rev. E {\bf 69}, 016116 (2004). }

\bibitem{ribot}  {\small  P.H. Chavanis, M. Ribot, C. Rosier and C. Sire,  Banach Center Publ. {\bf 66}, 103 (2004).  }

\bibitem{cll}  {\small  P.H. Chavanis, P. Lauren\c cot and M. Lemou, Physica A, {\bf 341}, 145 (2004).   }

\bibitem{naudts}  {\small Naudts, Physica A {\bf 332}, 279 (2004).}

\bibitem{tremaine}  {\small S. Tremaine, M. H\'enon and D. Lynden-Bell, Mon. Not. R. Astron. Soc. {\bf 219}, 285 (1986).}

\bibitem{csr}  {\small P.H. Chavanis, J. Sommeria and R. Robert, Astrophys. J. {\bf 471}, 385 (1996).}

\end{thebibliography}
\end{document}